\RequirePackage{lineno}
\documentclass[aps,prd,twocolumn,superscriptaddress,showpacs]{revtex4}

\usepackage{graphicx}
\usepackage{mycaption}

\newcommand{\ttb}{\ensuremath{t\bar{t}}}
\newcommand{\mttb}{\ensuremath{M_{t\bar{t}}}}
\newcommand{\ppb}{\ensuremath{p\bar{p}}}
\newcommand{\Et}{\ensuremath{E_T}}

\newcommand{\ba}{\begin{array}}
\newcommand{\ea}{\end{array}}
\newcommand{\bc}{\begin{center}}
\newcommand{\ec}{\end{center}}
\newcommand{\bn}{\begin{enumerate}}
\newcommand{\en}{\end{enumerate}}
\newcommand{\bq}{\begin{equation}}
\newcommand{\eq}{\end{equation}}
\newcommand{\bi}{\begin{itemize}}
\newcommand{\ei}{\end{itemize}}   
\newcommand{\bh}{\begin{math}}
\newcommand{\eh}{\end{math}}
\newcommand{\br}{\begin{flushright}}
\newcommand{\er}{\end{flushright}}
\newcommand{\bl}{\begin{flushleft}}
\newcommand{\el}{\end{flushleft}}
\newcommand{\bt}{\begin{tabular}}
\newcommand{\et}{\end{tabular}}
\newcommand{\flame}

\begin{document}

\begin{minipage}{\textwidth}
  \begin{flushright}
    FERMILAB-PUB-11-397-E\\
    CDF/PUB/TOP/CDFR/10473
  \end{flushright}
\end{minipage}
\vfill

\title{Search for resonant production of $t\bar{t}$ decaying to jets in $p\bar p$ collisions at $\sqrt{s}=1.96$ TeV}
\affiliation{Institute of Physics, Academia Sinica, Taipei, Taiwan 11529, Republic of China} 
\affiliation{Argonne National Laboratory, Argonne, Illinois 60439, USA} 
\affiliation{University of Athens, 157 71 Athens, Greece} 
\affiliation{Institut de Fisica d'Altes Energies, ICREA, Universitat Autonoma de Barcelona, E-08193, Bellaterra (Barcelona), Spain} 
\affiliation{Baylor University, Waco, Texas 76798, USA} 
\affiliation{Istituto Nazionale di Fisica Nucleare Bologna, $^{aa}$University of Bologna, I-40127 Bologna, Italy} 
\affiliation{University of California, Davis, Davis, California 95616, USA} 
\affiliation{University of California, Los Angeles, Los Angeles, California 90024, USA} 
\affiliation{Instituto de Fisica de Cantabria, CSIC-University of Cantabria, 39005 Santander, Spain} 
\affiliation{Carnegie Mellon University, Pittsburgh, Pennsylvania 15213, USA} 
\affiliation{Enrico Fermi Institute, University of Chicago, Chicago, Illinois 60637, USA}
\affiliation{Comenius University, 842 48 Bratislava, Slovakia; Institute of Experimental Physics, 040 01 Kosice, Slovakia} 
\affiliation{Joint Institute for Nuclear Research, RU-141980 Dubna, Russia} 
\affiliation{Duke University, Durham, North Carolina 27708, USA} 
\affiliation{Fermi National Accelerator Laboratory, Batavia, Illinois 60510, USA} 
\affiliation{University of Florida, Gainesville, Florida 32611, USA} 
\affiliation{Laboratori Nazionali di Frascati, Istituto Nazionale di Fisica Nucleare, I-00044 Frascati, Italy} 
\affiliation{University of Geneva, CH-1211 Geneva 4, Switzerland} 
\affiliation{Glasgow University, Glasgow G12 8QQ, United Kingdom} 
\affiliation{Harvard University, Cambridge, Massachusetts 02138, USA} 
\affiliation{Division of High Energy Physics, Department of Physics, University of Helsinki and Helsinki Institute of Physics, FIN-00014, Helsinki, Finland} 
\affiliation{University of Illinois, Urbana, Illinois 61801, USA} 
\affiliation{The Johns Hopkins University, Baltimore, Maryland 21218, USA} 
\affiliation{Institut f\"{u}r Experimentelle Kernphysik, Karlsruhe Institute of Technology, D-76131 Karlsruhe, Germany} 
\affiliation{Center for High Energy Physics: Kyungpook National University, Daegu 702-701, Korea; Seoul National University, Seoul 151-742, Korea; Sungkyunkwan University, Suwon 440-746, Korea; Korea Institute of Science and Technology Information, Daejeon 305-806, Korea; Chonnam National University, Gwangju 500-757, Korea; Chonbuk National University, Jeonju 561-756, Korea} 
\affiliation{Ernest Orlando Lawrence Berkeley National Laboratory, Berkeley, California 94720, USA} 
\affiliation{University of Liverpool, Liverpool L69 7ZE, United Kingdom} 
\affiliation{University College London, London WC1E 6BT, United Kingdom} 
\affiliation{Centro de Investigaciones Energeticas Medioambientales y Tecnologicas, E-28040 Madrid, Spain} 
\affiliation{Massachusetts Institute of Technology, Cambridge, Massachusetts 02139, USA} 
\affiliation{Institute of Particle Physics: McGill University, Montr\'{e}al, Qu\'{e}bec, Canada H3A~2T8; Simon Fraser University, Burnaby, British Columbia, Canada V5A~1S6; University of Toronto, Toronto, Ontario, Canada M5S~1A7; and TRIUMF, Vancouver, British Columbia, Canada V6T~2A3} 
\affiliation{University of Michigan, Ann Arbor, Michigan 48109, USA} 
\affiliation{Michigan State University, East Lansing, Michigan 48824, USA}
\affiliation{Institution for Theoretical and Experimental Physics, ITEP, Moscow 117259, Russia}
\affiliation{University of New Mexico, Albuquerque, New Mexico 87131, USA} 
\affiliation{Northwestern University, Evanston, Illinois 60208, USA} 
\affiliation{The Ohio State University, Columbus, Ohio 43210, USA} 
\affiliation{Okayama University, Okayama 700-8530, Japan} 
\affiliation{Osaka City University, Osaka 588, Japan} 
\affiliation{University of Oxford, Oxford OX1 3RH, United Kingdom} 
\affiliation{Istituto Nazionale di Fisica Nucleare, Sezione di Padova-Trento, $^{bb}$University of Padova, I-35131 Padova, Italy} 
\affiliation{LPNHE, Universite Pierre et Marie Curie/IN2P3-CNRS, UMR7585, Paris, F-75252 France} 
\affiliation{University of Pennsylvania, Philadelphia, Pennsylvania 19104, USA}
\affiliation{Istituto Nazionale di Fisica Nucleare Pisa, $^{cc}$University of Pisa, $^{dd}$University of Siena and $^{ee}$Scuola Normale Superiore, I-56127 Pisa, Italy} 
\affiliation{University of Pittsburgh, Pittsburgh, Pennsylvania 15260, USA} 
\affiliation{Purdue University, West Lafayette, Indiana 47907, USA} 
\affiliation{University of Rochester, Rochester, New York 14627, USA} 
\affiliation{The Rockefeller University, New York, New York 10065, USA} 
\affiliation{Istituto Nazionale di Fisica Nucleare, Sezione di Roma 1, $^{ff}$Sapienza Universit\`{a} di Roma, I-00185 Roma, Italy} 

\affiliation{Rutgers University, Piscataway, New Jersey 08855, USA} 
\affiliation{Texas A\&M University, College Station, Texas 77843, USA} 
\affiliation{Istituto Nazionale di Fisica Nucleare Trieste/Udine, I-34100 Trieste, $^{gg}$University of Udine, I-33100 Udine, Italy} 
\affiliation{University of Tsukuba, Tsukuba, Ibaraki 305, Japan} 
\affiliation{Tufts University, Medford, Massachusetts 02155, USA} 
\affiliation{University of Virginia, Charlottesville, Virginia 22906, USA}
\affiliation{Waseda University, Tokyo 169, Japan} 
\affiliation{Wayne State University, Detroit, Michigan 48201, USA} 
\affiliation{University of Wisconsin, Madison, Wisconsin 53706, USA} 
\affiliation{Yale University, New Haven, Connecticut 06520, USA} 
\author{T.~Aaltonen}
\affiliation{Division of High Energy Physics, Department of Physics, University of Helsinki and Helsinki Institute of Physics, FIN-00014, Helsinki, Finland}
\author{B.~\'{A}lvarez~Gonz\'{a}lez$^w$}
\affiliation{Instituto de Fisica de Cantabria, CSIC-University of Cantabria, 39005 Santander, Spain}
\author{S.~Amerio}
\affiliation{Istituto Nazionale di Fisica Nucleare, Sezione di Padova-Trento, $^{bb}$University of Padova, I-35131 Padova, Italy} 

\author{D.~Amidei}
\affiliation{University of Michigan, Ann Arbor, Michigan 48109, USA}
\author{A.~Anastassov}
\affiliation{Northwestern University, Evanston, Illinois 60208, USA}
\author{A.~Annovi}
\affiliation{Laboratori Nazionali di Frascati, Istituto Nazionale di Fisica Nucleare, I-00044 Frascati, Italy}
\author{J.~Antos}
\affiliation{Comenius University, 842 48 Bratislava, Slovakia; Institute of Experimental Physics, 040 01 Kosice, Slovakia}
\author{G.~Apollinari}
\affiliation{Fermi National Accelerator Laboratory, Batavia, Illinois 60510, USA}
\author{J.A.~Appel}
\affiliation{Fermi National Accelerator Laboratory, Batavia, Illinois 60510, USA}
\author{A.~Apresyan}
\affiliation{Purdue University, West Lafayette, Indiana 47907, USA}
% ...
\collaboration{CDF Collaboration\footnote{With visitors from $^a$Istituto Nazionale di Fisica Nucleare, Sezione di Cagliari, 09042 Monserrato (Cagliari), Italy,
$^b$University of CA Irvine, Irvine, CA  92697, USA,
$^c$University of CA Santa Barbara, Santa Barbara, CA 93106, USA,
$^d$University of CA Santa Cruz, Santa Cruz, CA  95064, USA,
$^e$CERN,CH-1211 Geneva, Switzerland,
$^f$Cornell University, Ithaca, NY  14853, USA, 
$^g$University of Cyprus, Nicosia CY-1678, Cyprus, 
$^h$Office of Science, U.S. Department of Energy, Washington, DC 20585, USA,
$^i$University College Dublin, Dublin 4, Ireland,
$^j$University of Fukui, Fukui City, Fukui Prefecture, Japan 910-0017,
$^k$Universidad Iberoamericana, Mexico D.F., Mexico,
$^l$Iowa State University, Ames, IA  50011, USA,
$^m$University of Iowa, Iowa City, IA  52242, USA,
$^n$Kinki University, Higashi-Osaka City, Japan 577-8502,
$^o$Kansas State University, Manhattan, KS 66506, USA,
$^p$University of Manchester, Manchester M13 9PL, United Kingdom,
$^q$Queen Mary, University of London, London, E1 4NS, United Kingdom,
$^r$University of Melbourne, Victoria 3010, Australia,
$^s$Muons, Inc., Batavia, IL 60510, USA,
$^t$Nagasaki Institute of Applied Science, Nagasaki, Japan, 
$^u$National Research Nuclear University, Moscow, Russia,
$^v$University of Notre Dame, Notre Dame, IN 46556, USA,
$^w$Universidad de Oviedo, E-33007 Oviedo, Spain, 
$^x$Texas Tech University, Lubbock, TX  79609, USA,
$^y$Universidad Tecnica Federico Santa Maria, 110v Valparaiso, Chile,
$^z$Yarmouk University, Irbid 211-63, Jordan,
$^{hh}$On leave from J.~Stefan Institute, Ljubljana, Slovenia, 
}}
\noaffiliation



\date{\today}

\begin{abstract}
This Letter reports a search for non standard model topquark
  resonances, $Z'$, decaying to $t\bar{t} \rightarrow W^+b W^-\bar{b}$
, where both $W$ decay to quarks.
We examine the top-antitop quark invariant mass spectrum for the presence of narrow
resonant states. The search uses a data sample of $\ppb$ collisions at a center of mass energy of 1.96
TeV collected by the CDF II detector at the Fermilab Tevatron,  with
an integrated luminosity of 2.8 fb$^{-1}$. No evidence
for top-antitop quark resonant production is found. We place upper
limits on the production cross section times branching ratio for a
specific topcolor assisted technicolor
model with width of $\Gamma_{Z'} = 0.012~M_{Z'}$. Within this model, we exclude $Z'$ boson
with masses below 805 GeV$/c^2$ at the 95\% confidence level.
\end{abstract}

\maketitle


The discovery of the top quark in 1995~\cite{TopDiscovery} completed the third generation of
quarks. After years of its discovery, the top quark plays an
important role in theoretical extensions of the standard model~(SM). Its large mass gives the top quark a special position
within the standard model and may shed light on the dynamics of
electroweak symmetry breaking. One SM extension, topcolor assisted
technicolor~\cite{Hill2}, predicts new strong dynamics which accounts for the spontaneous breaking of electroweak
symmetry and explains the large mass of the top quark. This model predicts a
vector particle ($Z'$), which couples primarily to the third
generation of quarks and has no significant couplings to leptons. The existence for
a narrow~($\Gamma_{Z'} = 0.012~M_{Z'}$) width $Z'$ resonance decaying to $\ttb$ pairs, using leptophobic
topcolor model~\cite{Harris} as the reference, has been probed since
the beginning of Tevatron operations both at CDF~\cite{CDFRunI} and
D0~\cite{D0RunI}. Other theories~\cite{Frederix} of physics beyond the SM predict heavy resonances that add a resonant part to the
SM $\ttb$ production mechanism. 

This Letter presents a search for narrow resonant states decaying to
top-antitop pairs. In the leptophobic topcolor model, top quarks decay as in the
SM via the weak interaction, nearly always to a $W$ boson and a $b$ quark. $W$ bosons decay into
  light fermion-antifermion pairs: a leptonic
decay~(32.4\%) into a charged lepton and a neutrino; or hadronic
decay~(67.6\%) into an up-type quark and a down-type quark.  All
previously reported searches have been analyses of top-antitop decays
in the lepton plus jets channel, where one of the $W$ bosons decays leptonically (to an
electron or a muon) and the other $W$ decays hadronically. 
This channel features a clean signature due
to the presence of a lepton in the  final state, and has a branching ratio
of 29\%. The result presented in this Letter is an analysis of $\ttb$
decays in the all-hadronic channel, where both $W$'s decay
hadronically.
 Because this topology features only multiple hadronic jets in the
 final state, it
 is subject to a considerable multijet QCD background.
We demonstrate that this background can be controlled and
significantly suppressed with a careful event selection. 
Analysis of $\ttb$ decays in the all-hadronic channel is
advantageous for several reasons: the channel offers the largest
branching ratio (46\%) of any of the $\ttb$ final states;  there is
no unobservable neutrino in the final state, which permits improved
resolution of the $\ttb$ invariant mass; finally, this sample is
orthogonal to that of previous analyses -- the result presented in
this Letter is complimentary to the previous results in the lepton plus jets channel. 
 
 $\ppb$ collision events analyzed in this paper were produced at the
 Tevatron collider at a center of mass energy of 1.96 TeV and
were recorded by the CDF II detector~\cite{CDFII}. The data sample
corresponds to a total integrated luminosity of 2.8 fb$^{-1}$. CDF II is a
general purpose particle detector. It consists of high precision tracking
systems for vertex and charged particle track reconstruction,
surrounded by electromagnetic and hadronic calorimeters for energy
measurement, and muon subsystems outside the calorimeter for muon detection. CDF II uses a cylindrical coordinate system with azimuthal
angle $\phi$, polar angle $\theta$ measured with respect to the
positive $z$ direction along the proton beam, and  the distance
$r$ from the beamline. The pseudorapidity, transverse
energy and momentum are defined as
$\eta=-\ln\left[\tan(\frac{\theta}{2})\right]$, $E_{t}=E\sin{\theta}$
and $P_{t}=P\sin{\theta}$, where $E$ and $P$ are the energy and momentum
of an incident particle.

The data were collected using a multijet on line event selection
~(trigger), which is implemented  in three stages. For
triggering purposes, the calorimeter is subdivided into a 24 $\times$ 24
grid of towers in $\eta$-$\phi$ space. 
At level 1, we require at least one trigger tower with transverse energy $E_{t}^{tow}$ $\geq$ 10~GeV. 
At level 2, we require the sum of the transverse energies of all the
trigger towers, $\sum E_{t}^{tow}$, to be $\geq$ 175 GeV and the presence of at least four clusters of trigger towers with $E_{t}^{cls}$ $\geq$ 15~GeV.
Finally, at level 3 we require four or more reconstructed jets with raw(uncorrected)
$E_{t}$ $\geq$~10 GeV, where jets are identified as clusters of energy depositions
in the calorimeter using a fixed cone~( $\Delta R = \sqrt{ (\Delta \phi)^2 +  (\Delta \eta)^2) } = 0.4$) algorithm~\cite{JetClu}.
The efficiency of this trigger selection on all-hadronic $\ttb$ events
is about 80$\%$. The main background present in this data sample is
QCD multijet production.

 The jet energies are corrected for calorimeter response, multiple interactions 
 and energy radiated outside the jet cone~\cite{nimjes}.
 Jets originating from a $b$ quark are identified by the {\sc secvtx} 
 ~\cite{secvtx} algorithm, which searches for tracks with non zero impact 
 parameter that result from the displaced decay of $B$ hadrons 
 inside the jet, and fits the tracks to a common vertex. If 
 this  vertex is significantly displaced from the primary interaction
 point, the jet is tagged as a $b$ jet. 

Events compatible with the signal are selected by requiring six or
seven jets with  $|\eta|<2$ and corrected $E_{t}>15$ GeV. To remove leptonic $t\bar{t}$ decays, we veto events with
well identified leptons~\cite{lep} or with significant imbalance in transverse momentum~\cite{metsig}. 
After all the preselections defined above, the SM $\ttb$ contribution
to the data sample is expected to be very small~(0.3\%). To enrich the
signal presence in the data sample we have to apply additional cuts, which we describe later in the paper.

The distinctive feature of this analysis is the use of likelihoods calculated by integrating
signal matrix elements both to perform $\ttb$ invariant mass reconstruction and to
suppress the overwhelming background. The full expression of the
likelihood for a given event with jet momenta
configuration $\vec{j}$ to be the result of SM $\ttb$ production and
decay is given by:
\begin{eqnarray}
\label{prob}
P(\vec{j}|m) =  \sum_{combi} \int \frac{dz_{a}dz_{b}f(z_{a})f(z_{b})}{4E_{a}E_{b}|v_{a}-v_{b}|} \prod_{i=1}^{6} \bigg[ \frac{d^{3}\vec{p}_{i}}{(2\pi)^{3}2E_{i}} \bigg] \nonumber \\
\label{eq:4-6} \times \frac{|\mathcal{M}(m,p)|^{2} (2\pi)^{4}\delta^{(4)}(E_{F}-E_{I}) TF(\vec{j}|\vec{p}) P_{t}(\vec{p})}{\sigma_{tot}(m)\epsilon(m)N_{combi}}
\end{eqnarray}, where $z_{a,b}$, $v_{a,b}$ and $E_{a,b}$ are the fractional momenta,
velocities and energies carried by partons $a$, $b$ and $f(z_{a,b})$ are the parton
distribution functions of colliding proton and antiproton, $\vec{j}$($\vec{p}$)
are jets (partons) four-momenta, $m$ is the top quark mass, $\mathcal{M}(m,p)$ is the SM $\ttb$
leading order matrix element, $\sigma_{tot(m)}\epsilon(m)$ is the SM
$\ttb$ production cross section times the selection efficiency both as
a function of $m$, and $E_{F}$($E_{I}$) is
a generic notation for the four-vector of the final (initial) state and
$P_{t}(\vec{p})$ for the transverse momentum of the $\ttb$ system. The
sum is performed over all jet to parton assignments
$N_{combi}$. The probability that a parton with energy $E_{p}$ is observed as a jet
with energy $E_{j}$ is given by the transfer function, $TF(\vec{j}|\vec{p})$, which
is parameterized as a function of parton energy and
pseudorapidity. Transfer functions are defined individually for the jets associated with $b$
and light jets, as they have different response in the
calorimeter. We construct the transfer functions from the events that have
all the jets uniquely matched to each individual parton within a
cone of $\Delta R = 0.4$. For each energy and $\eta$ region we use smoothed histogrammed
distributions of $1-E_{jet}/E_{parton}$ as  $TF(\vec{j}|\vec{p})$
parameterization. A sample of fully simulated $\ttb$ Monte Carlo~(MC) events,
generated using {\sc pythia} v6.2 ~\cite{pythia} with parton showering followed by the full simulation of 
the CDF\, II detector, and assuming $m$ = 175 GeV/$c^2$ is
used to obtain the $P_{t}(\vec{p})$ and $TF(\vec{j}|\vec{p})$
parametrizations.

The probability density, $P(\vec{j}|m)$, can be expressed with respect to any variable that is a function of parton four-momenta, in this case the
invariant mass of $\ttb$ pairs, $\mttb$. Integrating Eq.~(\ref{prob}) over
the parton variables times a delta function
$\delta(x-M_{t\bar{t}}(p))$ we obtain the probability density function for each event once the
jets are measured. We use the mean
value of this distribution as an estimator for $\mttb$.

To discriminate between SM background and $Z'$ signal events, we
calculate event quantities which are sensitive to the presence of a
signal and use them as inputs for a neural network which is trained to
separate the signal and the background. Keeping in mind that SM $\ttb$ is one of the
background samples for $Z'$ events, here we will refer to SM $\ttb$ as the signal
sample for the event selection purpose only. We train the neural network to
select events with the presence of $\ttb$ pairs and to veto dominant
QCD multijet production. Using SM $\ttb$ events as signal events to
optimize the event selection and enrich the $\ttb$ content of the sample accomplishes reasonable results for $Z'$
events as shown later in the paper. In addition, this
choice makes the search unbiased to a specific mass and the model of $Z'$
hypothesis used. 

A first set of 10 kinematic variables, summarized in Table\,~\ref{tab:nnvar}, has
already been shown to be  effective~\cite{ahprd} in reducing the QCD
background.  Significant distinguishing features of $\ttb$ production
in comparison to the QCD background are high $E_T$ jets, dijet resonances from $W$ decay and trijet
resonances from $t$ decay. The centrality is $C = \frac{\sum E_T}{\sqrt{\hat{s}}}$, where $\sqrt{\hat{s}}$ is 
the invariant mass of the multijet system. The aplanarity is defined as 
${A}=\frac{3}{2}{\cal{Q}}_1$, where ${\cal{Q}}_1$ is the smallest of the three normalized 
eigenvalues of the sphericity tensor, $M^{ab}= \sum _j P_{j}^a P_{j}^b$, calculated in the 
center of mass system of all jets, where $a$ and $b$ refer to the
spatial components of the jet four-momentum $P_j$.  
In Table~\ref{tab:nnvar}, $\theta^\star$ is a jet emission direction, represented by the angle between the jet direction, measured in the
center of mass frame of all jets, and the proton beam axis. For the
last variable, MED~(matrix element discriminant), we exploit the broad
set of information from
the event about its production and decay through the SM $\ttb$ matrix element. For each event we calculate `the minus log probability' of Eq.~(\ref{prob}) at 9
different top mass points, $m_t = 155, 160...195$ GeV/$c^2$, and use their sum as the final discriminator.

\begin{table}[htbp]

\caption{Neural network input variables.}\label{tab:nnvar}
\begin{center}
\begin{tabular}{c c}
\hline
\parbox[c][5mm][c]{15mm}{Variable} & Description \\
\hline
\parbox[c][5mm][c]{15mm}{ $ \sum \Et    $ }             & ~Scalar sum
of all jet $E_T$ \\
\parbox[c][5mm][c]{15mm}{ $ \sum _3\Et  $ }             & ~As above,
excluding two highest $E_T$ jets\\
\parbox[c][5mm][c]{15mm}{ $ C           $ }             & ~Centrality, 
defined in text\\
\parbox[c][5mm][c]{15mm}{ $ A           $ }             & ~Aplanarity,
defined in text\\
\parbox[c][5mm][c]{15mm}{ $ M_{2j}^{\mathrm{min}}$ }    & ~Minimum dijet invariant mass\\
\parbox[c][5mm][c]{15mm}{ $ M_{2j}^{\mathrm{max}}$ }    & ~Maximum dijet invariant mass\\
\parbox[c][5mm][c]{15mm}{ $ M_{3j}^{\mathrm{min}}$ }    & ~Minimum trijet invariant mass\\
\parbox[c][5mm][c]{15mm}{ $ M_{3j}^{\mathrm{max}}$ }    & ~Maximum trijet invariant mass\\
\parbox[c][5mm][c]{15mm}{ $ E_T^{\star, 1}$ }           &
~$E_T\sin^2\theta^\star$ for the highest $E_T$ jet\\
\parbox[c][5mm][c]{15mm}{ $ \langle E_T^\star\rangle$ } &  ~Geometric
mean of $E_T$ of remaining $N-2$ jets\\
\parbox[c][5mm][c]{15mm}{ MED } & ~Constructed from matrix element\\
\hline
\end{tabular}
\end{center}
\end{table}

\begin{figure}[htbp]
  \begin{center}
    \includegraphics[width=3.4in]{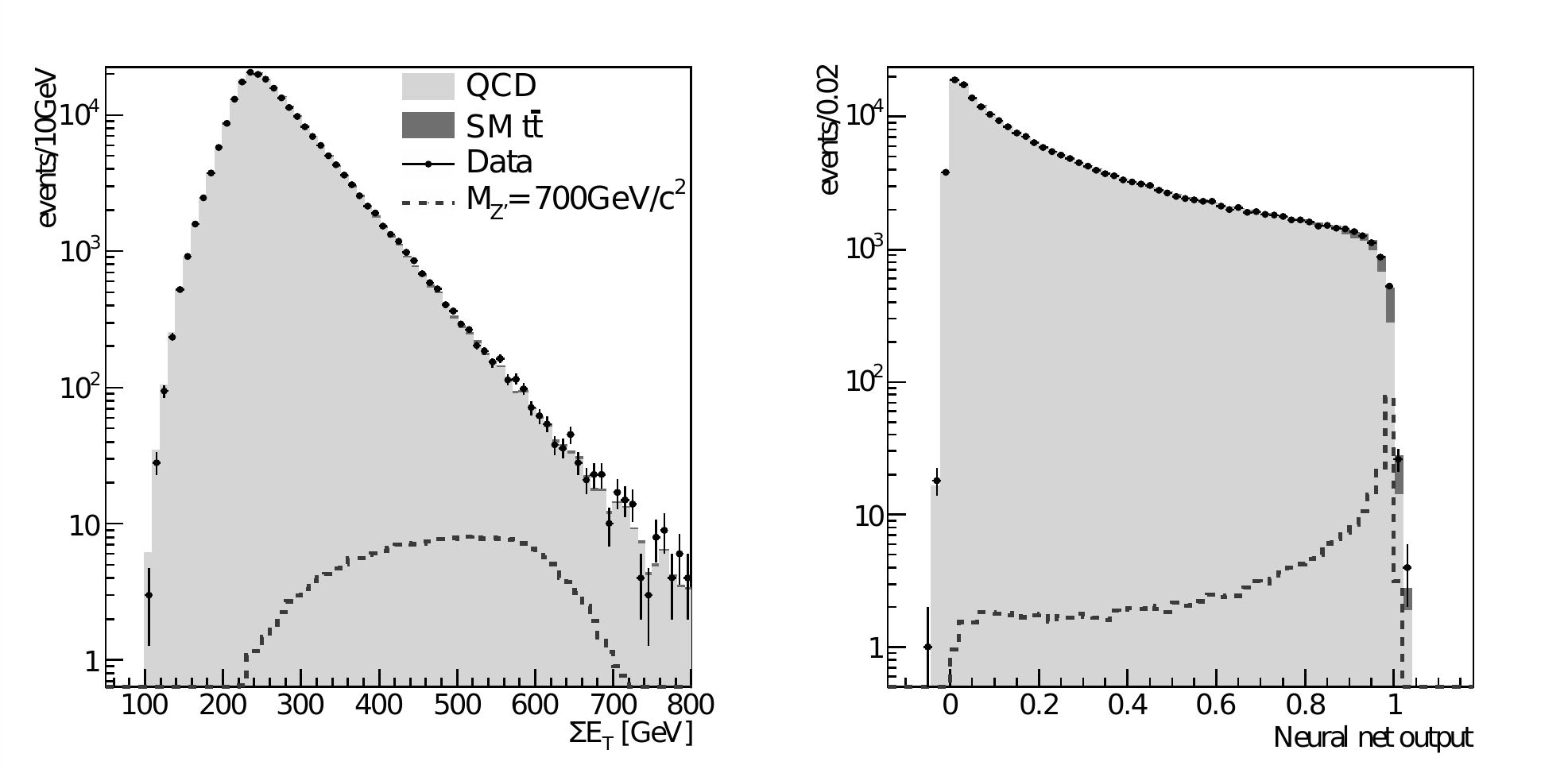}
    \caption[Neural net input variables and the output node value]{ $\sum
      \Et$ and neural net output value for QCD dominated
      data~(black points), SM $\ttb$~(dark grey histogram), QCD
      prediction~(light grey histogram) and 700 GeV/$c^2$ $\ttb$
      resonance normalized to the expectation~(dashed histogram).}
    \label{fig:NNVar2}
  \end{center}
\end{figure}

Having defined the variables, to separate $\ttb$ from background
events, we use them as inputs to a neural network~\cite{mlp} with two hidden
layers and one output node. The neural network is trained on samples of signal and background events
 with $6\le N_{\rm jets}\le 7$. To model the signal events we use the
 {\sc pythia} MC generator at leading order(LO) to produce SM
 $\ttb$ events assuming a top quark mass of
 $m_{\mathrm{top}}=175$\,GeV/$c^2$ and the theoretical cross section of 6.7
pb~\cite{theo-xsec}. We use the multijet data 
events as the background sample for training the neutral net since the $\ttb$ contribution is expected to be negligible. After the training the value of the output node,
$NN_{\mathrm{out}}$, is used as a discriminator between
 signal and background. Its distribution is shown in Fig.~\ref{fig:NNVar2}. In
 addition, we show the comparison of the QCD dominated
 data, MC generated SM $\ttb$ and MC generated $Z'$ events for one of the input variables  $\sum \Et$.

In the final event selection we require a cut on the neural net
output, $NN_{\mathrm{out}}>$0.93, and at least one jet tagged as
having originated from a $b$ quark. The neural net
requirement was optimized to	suppress the QCD background while
enhancing the content of $\ttb$ events by maximizing the SM $\ttb$
significance. Table~\ref{tab:accept} shows the selection
efficiencies for SM $\ttb$ and $Z'$ events
after final event selection cut. There are 2086 events surviving these
final selection criteria including 680 SM $\ttb$ events as estimated from the simulated event
sample and assuming the NLO theoretical cross section~\cite{theo-xsec}. The
remaining events are from QCD multijet processes plus a potential
signal contribution from $Z'$ events.

\begin{table}[htbp]
\caption{Table of cross sections, $\sigma$, and acceptances, $\epsilon \pm \delta \epsilon$(tot.), for $Z'$ and SM $\ttb$ events.}\label{tab:accept}
\begin{center}
\begin{tabular}{c c c}
\hline
$M_{Z'}$(GeV/$c^2$)  & $\sigma$[pb] & $\epsilon \pm \delta \epsilon$ \\
\hline
SM $\ttb$ & 6.7 &  3.8 $\pm$ 0.5 \\
450 &  8.96 & 4.2 $\pm$ 0.5  \\
500 &  5.66 & 4.7 $\pm$ 0.5  \\
550 &  3.40 & 5.3 $\pm$ 0.5  \\
600 &  2.09 & 5.7 $\pm$ 0.5  \\
650 &  1.31 & 5.8 $\pm$ 0.4  \\
700 &  0.78 & 5.6 $\pm$ 0.4  \\
750 &  0.47 & 5.2 $\pm$ 0.3  \\
800 &  0.28 & 4.6 $\pm$ 0.3  \\
850 &  0.16 & 4.0 $\pm$ 0.2  \\
900 &  0.10 & 3.6 $\pm$ 0.2  \\
\hline
\end{tabular}
\end{center}
\end{table}

\begin{figure}[htbp]
  \begin{center}
    \includegraphics[width=3.5in]{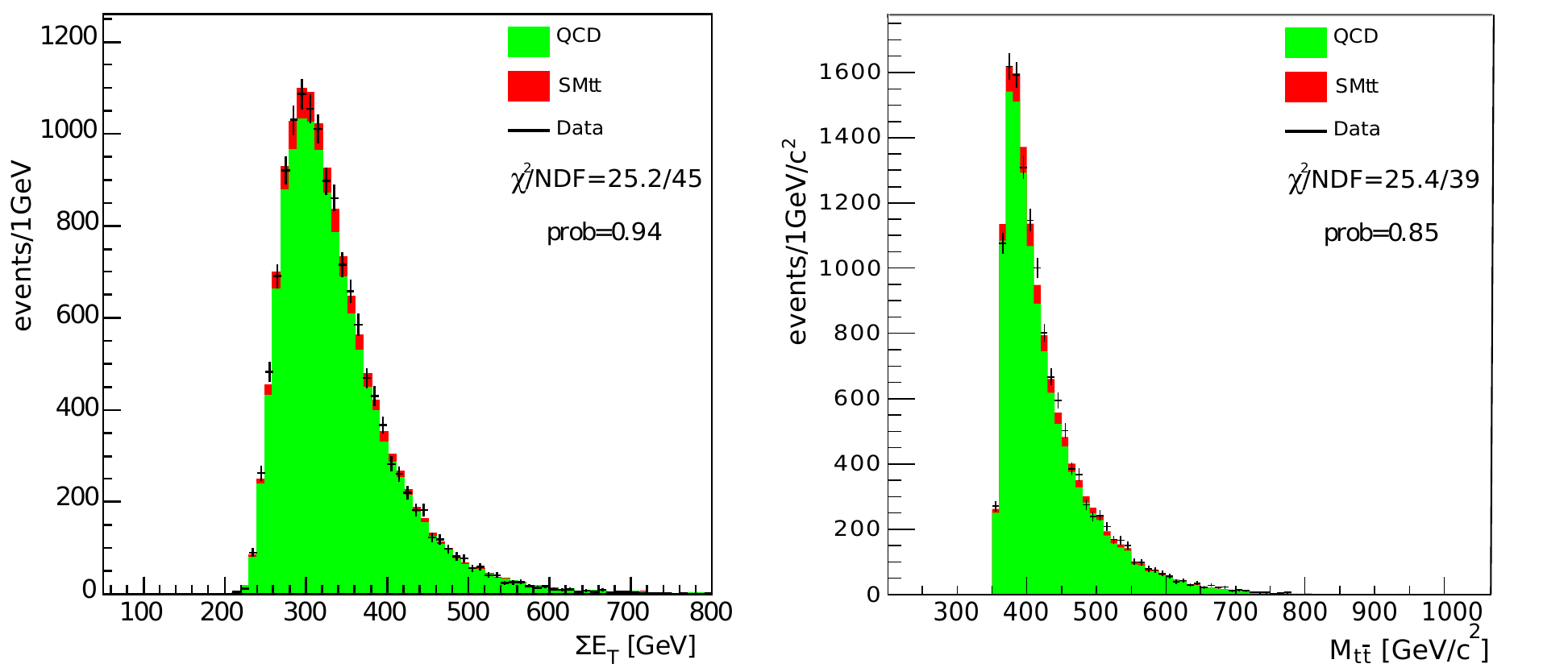}
    \caption[]{Distributions of $\sum
      \Et$ and $\mttb$ in the control region $ 0.75 < NN_{\mathrm{out}} \le 0.93$.}
    \label{fig:CR}
  \end{center}
\end{figure}

The dominant background is multijet production via QCD, where one of
the $b$ jets can originate from heavy flavor~($b$ or $c$) quark pair
production or from misidentified light flavor quark jets.
Due to the large theoretical uncertainties on the production
cross section, we use a data-driven approach to estimate the QCD background. From a data sample with 4 or 5
jets, which is overwhelmingly from QCD production (SM $\ttb$ fraction
less than $5\cdot10^{-4}$), we build a {\it tag
  rate} matrix. In this procedure, we
parametrize the probability for each jet to be identified as a $b$
jet. The parametrization includes the dependence on the
transverse energy of the jet, the number of tracks associated to the jet,
and the number of reconstructed collision vertices in the event. Once we define the
probability for a single jet to be tagged, we can use the tag rate
matrix to estimate the probability for an event to have one or
two $b$-tagged jets. The tag rate matrix
is applied to 6 or 7 jet data events before the $b$-tagging requirement to
predict the QCD background for events in the final selected sample. To test our background model, we consider
several control regions, defined by the neural net output value
$NN_{\mathrm{out}} \le 0.25$, $ 0.25 < NN_{\mathrm{out}} \le 0.75$, $ 0.75 < NN_{\mathrm{out}} \le 0.93$. For all the regions we
find a very good agreement between the model and the observed data. Figure~\ref{fig:CR}
shows the distributions of $\sum \Et$ and $\mttb$ in the control region $ 0.75 <
NN_{\mathrm{out}} \le 0.93$.

To measure and set the confidence level intervals on resonant $\ttb$ production given the observed
$\mttb$ spectrum, we start with the following likelihood,
\begin{eqnarray}
\label{like} 
L(\vec{n}|\sigma, \vec{\nu})= \prod_{i}
e^{-\mu_i}\frac{\mu_i^{n_i}}{n_i !}
\end{eqnarray}
which is the prior probability of observing $\vec{n}$, where $n_i$ is
the observed number of events in the ${\it  i}$th bin of
the $\mttb$ distribution, and $\mu_i$ is the expected number of events
in the same bin and is given by $\mu_i = \sigma_s (A_s -
A_s^{cont})T_s^i + \sigma_{tt} (A_{tt}-A_{tt}^{cont})T_{tt}^i +
N_{QCD}T_{QCD}^i$, which depends on the assumed signal cross section,
$\sigma_s$, the SM $\ttb$ cross section, $\sigma_{tt}$, the signal and SM $\ttb$ effective
acceptances, $A_s, A_{tt}$, the number of expected QCD events,
$N_{QCD}$, and the fraction of events in ${\it i}$th $\mttb$ bin, $T_{s}^i,
T_{tt}^i, T_{QCD}^i$ for the signal, SM $\ttb$ and QCD distributions,
respectively. As the QCD background prediction is performed using the data
sample itself, the presence of SM $\ttb$ and
assumed signal events must be subtracted from QCD background
estimation. We calculate the number of residual contamination events
in QCD background prediction, using assumed
and theoretical cross sections for signal and SM $\ttb$ events, and
their effective acceptances for the residual contamination terms
$A_s^{cont}$ and $A_{tt}^{cont}$ by applying the tag rate matrix to the signal and SM $\ttb$
samples before the tagging requirement, and 

 We use Bayes' theorem to connect the likelihood of the measurement to
 the posterior probability density, which is used to set the upper limits.

\begin{eqnarray}
\label{post}
p(\sigma|\vec{n})  = \int d\vec{\nu}\, p(\sigma, \vec{\nu}| \vec{n}) = \int d\vec{\nu}\, L (\vec{n}|\sigma, \vec{\nu}) \pi (\sigma,\vec{\nu}) /p(\vec{n})
\end{eqnarray}
where $\pi(\sigma,\vec{\nu})$ is the prior probability density, and
$p(\vec{n})=\int d\vec{\nu} \int d\sigma L (\sigma,\vec{\nu}|\vec{n})
\pi (\sigma,\vec{\nu})$.

There are two types of uncertainties we have to consider. The first
type does not change the shape of the $\mttb$ distribution but only
the acceptances. The second type
affects both the shape and normalization; we'll refer as shape uncertainties. Uncertainties that do
not change the shape of the templates (distributions) are
incorporated as nuisance parameters and integrated over in
Eq.~(\ref{post}). In this respect, Eq.~(\ref{post}) includes not only the statistical uncertainty of the data, but also
the source of systematic uncertainties on: signal and
SM $\ttb$ acceptances, SM $\ttb$ cross section, QCD normalization and
integrated luminosity.

One of the shape uncertainties we consider is the  jet energy scale
corrections. After the jet energy corrections we are left with an uncertainty on
the jet energy scale. A change in the jet energy scale modifies both
the acceptances and the template shapes. To
account for shape uncertainties and jet energy scale in particular, we generate a set of
pseudoexperiments using the shifted templates and acceptances. This results in a shifted
reconstructed cross section with respect to the nominal one. The mapping
of this shift versus the input cross section provides an evaluation of
the impact of the jet energy scale uncertainty at any given
cross section. The complete list of the shape uncertainties sources
we considered are jet energy scale, initial and the final state
radiation, and uncertainty on proton and antiproton parton distribution functions.
Assuming that the nature of shape uncertainties follow
normal distribution, we convolute the posterior probabilities with a
Gaussian whose width is equal to the quadrature sum of
individual shape uncertainties.

After including the shape uncertainties, the posterior density function is used to define the upper and lower limits at any given confidence
level (C.L.). If the lower limit is zero then the data is considered consistent with the SM at that level of confidence. 
We also extract the reconstructed cross section as the most probable
value of the posterior. To obtain the sensitivity of the reconstruction
algorithm we generate 1000 simulated experiments in the signal
null hypothesis and extract the 95\% C.L. expected upper limit, defined as the median of the upper limits distribution. 
This entire exercise is repeated for various resonance masses from 450
GeV/$c^2$ to 900 GeV/$c^2$. Together with the theoretical cross section versus mass curves these limits are used to
exclude certain mass ranges.

\begin{figure}[htbp]
  \begin{center}
    \includegraphics[width=2.8in]{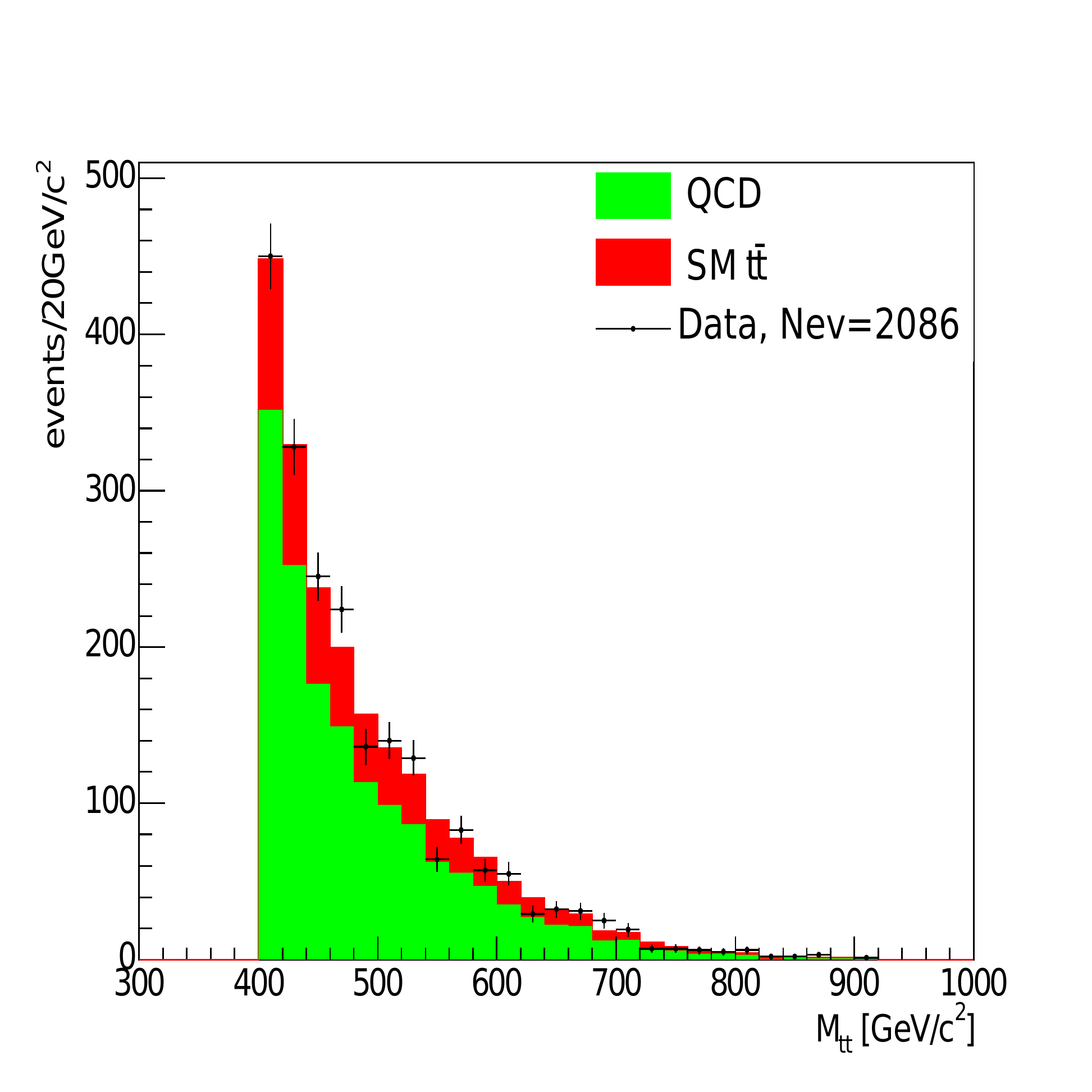}
    \caption{Reconstructed $\mttb$ vs the SM expectation in the search region above the 400 GeV/$c^2$ cut.}
    \label{fig:Data}
  \end{center}
\end{figure}

\begin{figure}[htbp]
  \begin{center}
    \includegraphics[width=2.7in]{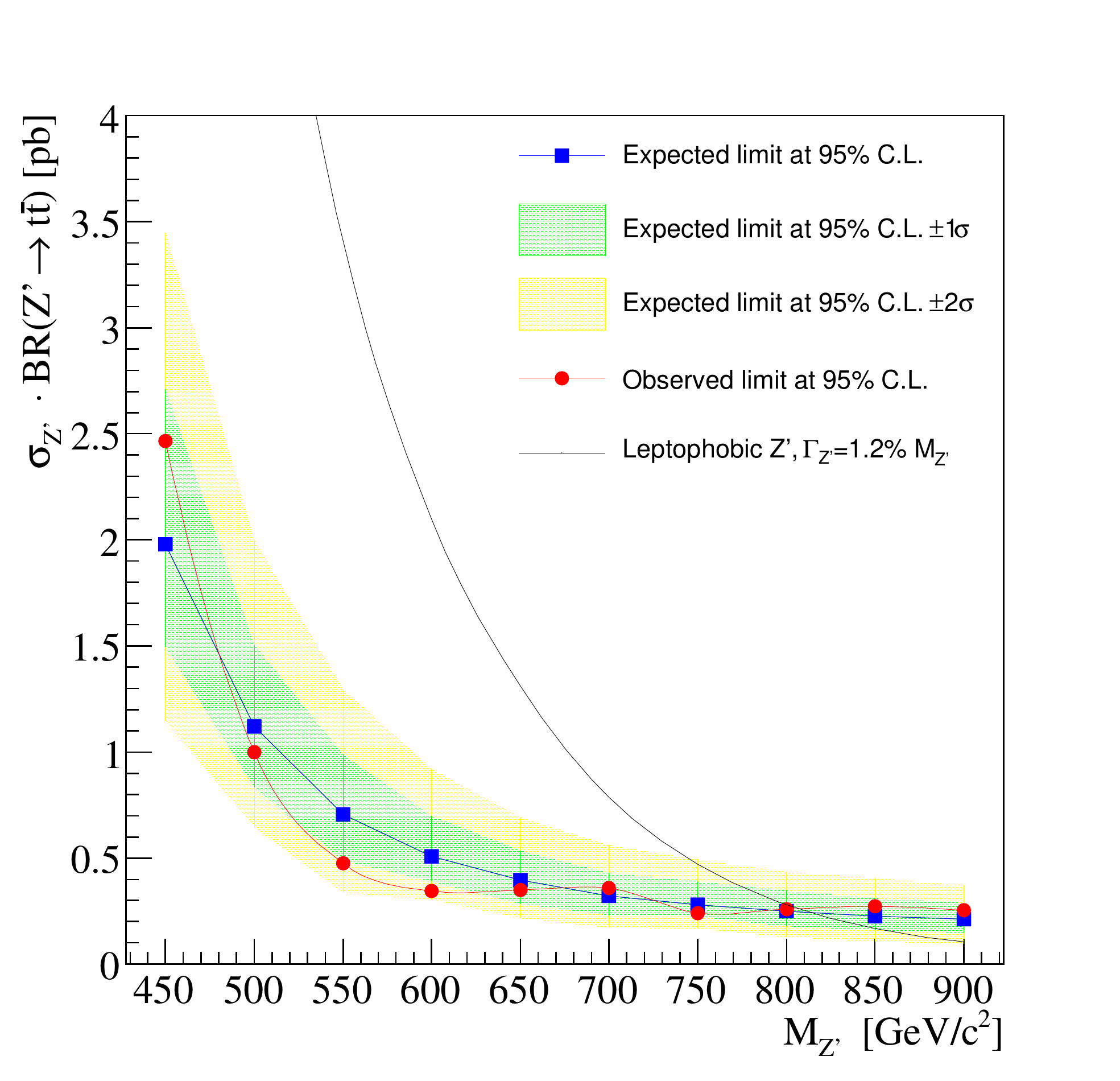}
    \caption{
      Expected and observed upper limits on leptophobic topcolor
      $Z'$ in 2.8 fb$^{-1}$ of CDF II data. The blue line is the median expected
      upper limit with the assumption of no signal, the red line is the observed
      limit and the black line is the cross section prediction for leptophobic topcolor
      $Z'$ production.}
    \label{fig:Brazil}
  \end{center}
\end{figure}

In this analysis we consider the data gathered by CDF II between 2002-2008 at the Tevatron.
The $\mttb$ distribution for the 2086 events surviving the final event
selection criteria is shown in Fig.~\ref{fig:Data} and is consistent
with the SM expectations. The resulting 95\% C.L. upper limits on
$\sigma(p\bar p \rightarrow Z')\cdot BR(Z' \rightarrow t\bar t )$ as a
function of $\mttb$ are shown in Fig.~\ref{fig:Brazil} together with
expected limits derived from pseudoexperiments that include the
SM background hypothesis only. These limits can be used to exclude a
leptophobic topcolor resonance candidates with a mass less than 805
GeV/$c^2$ at 95\% C.L., assuming the width of the resonance is $\Gamma_{Z'} = 0.012~M_{Z'}$. The previous searches were performed in
the lepton plus jets channel only, and  the most recent results were
conducted by CDF II~\cite{ZprimeCDF} and
D0~\cite{ZprimeD0}. Using Tevatron data corresponding to 1 fb$^{-1}$ and 0.9
fb$^{-1}$ integrated luminosity respectively, they found no evidence
for $\ttb$ resonant production. For the same benchmark model of
leptophobic topcolor $Z'$, the upper limits were set at 720 GeV/$c^2$ and
700 GeV/$c^2$ for CDF II and D0, respectively. 

In conclusion, we have performed a search for a heavy resonance
decaying into $\ttb$  using data with 2.8 fb$^{-1}$ integrated luminosity
in the all-jets channel. 
No evidence is observed and we set upper limits on the production
cross section times branching ratio at the 95\% C.L.  For one leptophobic topcolor production mechanism, we exclude 
masses up to 805 GeV/$c^2$.




We thank the Fermilab staff and the technical staffs of the participating institutions for their vital contributions. This work was supported by the U.S. Department of Energy and National Science Foundation; the Italian Istituto Nazionale di Fisica Nucleare; the Ministry of Education, Culture, Sports, Science and Technology of Japan; the Natural Sciences and Engineering Research Council of Canada; the National Science Council of the Republic of China; the Swiss National Science Foundation; the A.P. Sloan Foundation; the Bundesministerium f\"ur Bildung und Forschung, Germany; the Korean World Class University Program, the National Research Foundation of Korea; the Science and Technology Facilities Council and the Royal Society, UK; the Institut National de Physique Nucleaire et Physique des Particules/CNRS; the Russian Foundation for Basic Research; the Ministerio de Ciencia e Innovaci\'{o}n, and Programa Consolider-Ingenio 2010, Spain; the Slovak R\&D Agency; the Academy of Finland; and the Australian Research Council (ARC).


\begin{references}

\bibitem{TopDiscovery}
  F.~Abe {\it et al.}  (CDF Collaboration),
  Phys.\ Rev.\ Lett.\  {\bf 74}, 2626 (1995);
	S.~Abachi {\it et al.}  (D0 ~Collaboration),
  ibid., 2632 (1995).

\bibitem{Hill2} 
C.T. Hill,  Phys. Lett. B $\mathbf{345}$, 483 (1995).

\bibitem{Harris} 
R.M. Harris, C.T. Hill, and S.J. Parke, 
arXiv:hep-ph/9911288, (1999).

\bibitem{CDFRunI} T.\, Affolder {\it et al.} (CDF Collaboration),
 Phys.\ Rev.\ Lett.\  {\bf 85}, 2062 (2000);


\bibitem{D0RunI}  V.\, Abazov  {\it et al.} (D0 Collaboration), 
Phys.\ Rev.\ Lett.\  {\bf 92}, 221801 (2004)

\bibitem{Frederix} R.~Frederix and  F.~Maltoni, JHEP {\bf 0901}, 047 (2009).

\bibitem{CDFII} 
D.~Acosta {\it et al.}  (CDF Collaboration),
		  Phys.\ Rev.\  D {\bf 71}, 032001 (2005).

\bibitem{JetClu} F.~Abe {\it et al.}, Phys.~Rev.~D {\bf 45}, 1448
  (1992).


\bibitem{nimjes} A. Bhatti {\it et al.} (CDF Collaboration), Nucl. Instrum. Methods Phys. Res., Sect. A {\bf 566}, 2 (2006).



\bibitem{secvtx}D.~Acosta~{\it et al.} (CDF Collaboration),
  Phys. Rev. D {\bf 71}, 052003 (2005).



\bibitem{lep}  A. Abulencia {\it et al.} (CDF collaboration), Phys. Rev. D
  {\bf 74}, 072006(2006)

\bibitem{metsig} D. Acosta {\it et al.} (CDF Collaboration),
  Phys. Rev. Lett. {\bf 96}, 202002 (2006).


\bibitem{pythia} T. Sjostrand {\it et al.}, Computer Physics
  Commun. {\bf 135} (2001) 238.

\bibitem{ahprd}  T.\,Aaltonen {\it et al.} (CDF Collaboration), Phys. Rev. D {\bf 76}, 072009 (2007).


\bibitem{mlp} MLPFIT: A Tool for Multi-Layer Perceptron, \\
http://schwind.home.cern.ch/schwind/MLPfit.html

\bibitem{theo-xsec} M.\,Cacciari {\it et al.}, J.\,High\,Energy\,Phys. 09 (2008) 127; 
                    N.\,Kidonakis and R.\,Vogt, Phys.\,Rev. D {\bf 78}, 074005 (2008);
                    S.\,Moch and P.\,Uwer,
                    Nucl.\,Phys.\,Proc.\,Suppl. 183:75 (2008).


\bibitem{ZprimeCDF}  T.\,Aaltonen {\it et al.} (CDF Collaboration),
  Phys. Rev. D {\bf 77}, 051102 (2008)

\bibitem{ZprimeD0}  V.M. Abazov {\it et al.} (D0 Collaboration),
  Phys. \ Lett. \ B {\bf 668}, 98 (2008)

\end{references}
\end{document}